\documentclass[
    ,final            
  ]
  {aipproc}

\layoutstyle{8x11double}
\begin{document}

\title{Algorithms for lattice QCD: progress and challenges}


\author{Stefan Schaefer}{
  address={CERN, Physics Department, CH-1211 Geneva 23, Switzerland},
  altaddress={Humboldt Universit\"at zu Berlin, Institut f\"ur
Physik, Newtonstr. 15, 12489 Berlin, Germany}
}

\begin{abstract}
The development of improved algorithms for QCD on the
lattice has enabled us to do calculations
at small quark masses and get control over the chiral
extrapolation. Also finer lattices have become possible,
however, a severe slowing down associated with
the topology of the gauge fields has been observed.
This may prevent simulations of lattices fine enough
for controlling the continuum extrapolation.
This conference contribution introduces the basic concepts behind
contemporary lattice algorithms, the current knowledge
about their slowing down towards the continuum and its
consequences for future lattice simulations.
\mbox{}\hfill{\tt CERN-PH-TH/2010-274, SFB/CPP-10-120}
\end{abstract}

\maketitle


\section{Introduction}

Lattice QCD has witnessed a dramatic progress during the last decade.
Whereas simulations ten years ago where still carried out either in the
quenched approximation or at sea quark masses far above the physical
values, we can now calculate with much lighter quarks  with masses down
to the physical values of the up and down quark. Only part of this
progress is due to increased computer resources. The larger part comes
from improved algorithms,  in particular the way the fermion
determinant, responsible for the effects of the sea quarks, is included
in the simulations.  The ideas behind this progress will be reviewed in
the first part of this contribution.

By their nature, lattice calculations are never done ``at the physical
point'', i.e. in  continuum space-time, infinite volume and with six
dynamical quark flavors at precisely the physical masses, ideally 
taking the effect of the full standard model into account. In particular, a
lattice simulation will always be at a finite lattice spacing and the
result at several {\sl fine} lattice spacings $a$ has then to be
extrapolated to $a=0$. The systematic error associated with this
extrapolation depends on the ability to simulate at lattice spacings
which are much smaller than the scales involved in the problem. For the
physics of light quarks $\Lambda_\mathrm{QCD}$ is the relevant scale.
For relativistic heavy quarks, however, additionally the  scale set by
the quark's mass plays an important role and the lattice spacing $a$ has
to be sufficiently small, i.e. $am_q\ll1$. For the charm quark, e.g., 
one therefore needs lattice spacings well below 0.05fm for precision 
results.

In the generation of the ensembles fine enough to control the continuum
extrapolation, however, a grave problem occurs: some modes in the
simulation move increasingly slowly, in particular the phenomenon is
visible in the topological charge of the gauge configuration. The
simulation rarely tunnels between topological sectors.  

Some critical slowing down is expected in any simulation as critical
points are approached, here the continuum limit.  A typical rate of this
increase is with the second power of the correlation length. In gauge
theory, however, we find a critical exponent of $z\approx5$ instead of
two, making simulations of fine lattices practically impossible.  The
details of these statements are covered in the second half of this writeup.
The writeup finishes with an overview of the current state of the
lattice simulations with Wilson fermions.  The status of staggered quark
simulations has been covered by Gottlieb at this conference\cite{gottliebQC}.

\section{Dynamical fermions}

Because of their anti-commuting nature, there is no natural way to treat
fermions in numerical simulations.  The textbook version of the path
integral for QCD is unfortunately not suitable for numerical simulations
since it contains integrals over Grassmann variables.  They can be
performed analytically, which introduces the determinant of the Dirac
operator, with the remaining integrals over the gauge degrees of freedom
$U$ 
\begin{equation}
Z=\int [d U]\, \prod_f^{N_f} \det[ D(m_f)]\,  \exp(-S_g[U])
\label{eq:1}
\end{equation}
with $S_g$ the gauge action and $D(m_f)$ the lattice version of the
Dirac operator with quark mass $m_f$.  Since $D$ is a large matrix,
computing its determinant is virtually impossible and it is replaced by
a path integral over bosonic ``pseudo-fermion'' fields $\phi$
\begin{equation}
\det D^2(m_f) \propto \int [d\phi][d\phi^\dagger] \exp(-\phi^\dagger 
\frac{1}{D^\dagger D} \phi)\ ,
\label{eq:2}
\end{equation}
which works for even numbers of degenerate flavors, because the matrix
in the exponent has to be Hermitian positive definite.  Although
applying this identity seems innocent, using it in a straight forward
manner simulating the resulting  path integral (containing gauge and
pseudo-fermion fields) with the Hybrid Monte Carlo  (HMC)
algorithm\cite{Duane:1987de} turns out to be unfeasible for light quarks
and fine lattices. This was famously summarized by Ukawa at the Lattice
conference in 2001\cite{Ukawa:2002pc}, where he gave as the cost of
generating a decent sized ensemble for two-flavors of dynamical Wilson quarks as
\begin{equation}
C\left[\frac{n_\mathrm{conf}}{1000} \right]
\left[\frac{m_\pi/m_\rho}{0.6} \right]^{-6}
\left[\frac{L}{3\mathrm{fm}} \right]^{5}
\left[\frac{0.1\mathrm{fm}}{a} \right]^{7}
\label{eq:bw}
\end{equation}
with $C=2.8~\mathrm{Tflops~years}$. At the physical value of the pion mass, 
where $m_\pi/m_\rho\approx0.17$, this would be impossible even with today's machines.
Just physically light quarks, for which 3fm are  too small, would require
a peta-flops machine for several years despite the coarse lattice spacing of 0.1fm.

There are two, related, basic insights which lead to the progress of
the last decade.  The first was that the estimate of the determinant,
which is provided by {\sl one} realization of the pseudo-fermion field
$\phi$, is not good enough.  Better estimators have to be used, but in a
way, which is meaningful from the physics point of view and which also
can be introduced into lattice QCD algorithms.

The second insight, which often also provides a solution to the first,
is that the ultra-violet and the infra-red physics of the theory are
different and also need to be treated differently. By separating the
two, one can deal with them according to their requirements. But the
successful methods to achieve this splitting  also provide improved
estimators of the fermion determinant.

The initial break-through into this direction is by Hasenbusch with his
mass preconditioning\cite{algo:GHMC,algo:GHMC3}, where first, the
fermion matrix is split into two parts
\begin{equation}
\det D(m_f) = \det D(M) \det\big[ D(m_f) D^{-1}(M)\big]
\label{eq:mp}
\end{equation}
with $M$ a mass larger than $m_f$. The first term is therefore
dominated by the UV physics, whereas the second part is largely
infrared. On each of the two terms, Eq.~\ref{eq:2} is applied and
the resulting partition function can be simulated with the standard
HMC algorithm, yielding a dramatic speed-up, for the right choice
of $M$, of course.

\begin{figure}[t]
\includegraphics[width=0.27\textwidth]{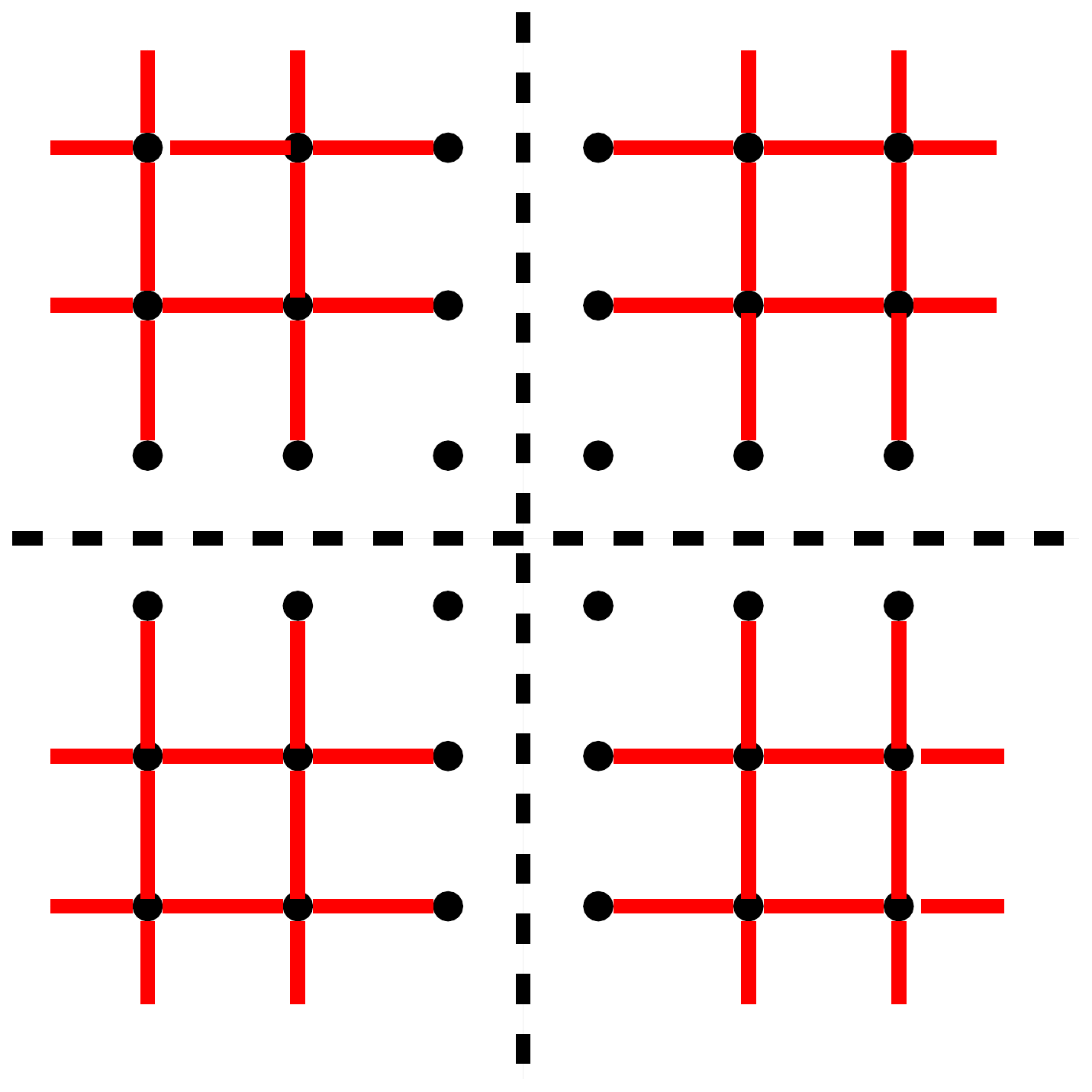}
\caption{\label{f:dd} In the DD-HMC algorithm, the lattice is split into
blocks, thereby separating the UV from the IR part of the fermion action.}
\end{figure}

Another successful split-up of the determinant is domain decomposition
(giving the DD-HMC its name), which was introduced by L\"uscher in
Ref.~\cite{algo:L2}, see Fig.~\ref{f:dd} for an illustration.  Here, the
lattice is decomposed into blocks and the Dirac operator is split into
one living only inside the blocks and a correction term, which accounts
for the rest
\[
\det D(m_f) = \prod_\mathrm{blocks} \det D_\mathrm{block}(m_f) \cdot \det R
\]
This provides an obvious geometric separation in  ultra-violet and infra-red 
part with the corresponding speed-up in the algorithm.

And finally, the third way of splitting the determinant leading to the  RHMC\cite{algo:RHMC}
is to use
\[
\det D = \prod_{i=1}^N \det D^{1/N}  \ .
\]
The separation between IR and UV is less clear, however, it has the advantage
that the fermion flavors do not have to come in even numbers of degenerate quarks.
It is therefore the method of choice for simulating the strange and charm quark, using
the identity $\det D = \det \sqrt{D^\dagger D}$.

The algorithms profit from the separation of the UV and the IR because
these two contributions evolve on very different time scales. The gauge
field has much faster fluctuations than most of the UV of the fermions,
which in turn evolve much faster than the infra-red. This makes the use
of specialized methods
possible\cite{Sexton1992665,AliKhan:2003mu,algo:urbach}, which move the
modes on separate time scales.

For example, just the block decomposition of the DD-HMC algorithm
leads to a cost formula of
\begin{equation}
C\left[\frac{n_\mathrm{conf}}{1000} \right]
 \left [ \frac{ 20\mathrm{MeV}}{m} \right ]
\left [ \frac{L}{3\mathrm{fm}} \right ]^5
\left [ \frac{0.1 \mathrm{fm}}{a} \right ]^6 \ , 
\label{eq:costdd}
\end{equation}
with $C=0.5$ Tflops~years and $m$ the running $\overline{MS}$ sea quark
mass at $2$GeV \cite{cern:I}. To compare with Eq.~\ref{eq:bw}, we note
that to leading order in the chiral expansion $m_\pi^2 \propto m$. Thus
the exponent governing the cost of going chiral, which was 6 ten years
ago, has dropped to 2. Equally important, the overall normalization is
reduced by a factor of 100.  With mass preconditioning, similar
performance can be reached\cite{Marinkovic:2010eg}.

To summarize, of the three algorithms, the RHMC follows the most the idea of 
an improved estimator of the fermion determinant. It provides a splitting
of the determinant into equal parts, reducing the fluctuations introduced
by the pseudo-fermions. The separation of IR and UV, however, is not as obvious,
but both profit from the improved estimation.

At the other end, the block decomposition of DD-HMC is clearly designed for 
the separation of the UV from the IR.
Since the action is split into the two parts, whose relative size can be
tuned by the size of the blocks, also a better estimator is provided.

The mass preconditioning takes an intermediate position. The IR/UV
separation is softer, since in both factors of Eq.~\ref{eq:mp} contain
both parts of the spectrum, however, with  different weights. But it
also provides a clear handle on getting a better estimator for the
fermion determinant. The identity of Eq.~\ref{eq:mp} can be iterated and
with a larger number of suitably chosen masses $M_i$, a systematically
better estimate can be reached.

\subsection{Solver}
All evidence speaks very much in favor of the efficiency of the simple
observation that the infra-red is different from the ultra-violet part of the 
physics and that respecting this physics can lead to this very beneficial
effects. The art is to bring it in a feasible and  cost-efficient way to the 
numerical simulations.

It has long been known that the infra-red part of the Dirac operator is
responsible for the high cost of solving the Dirac equation, which is
the most expensive part of lattice simulations.  In some special
applications, therefore, the low eigenvectors are removed, leading to a
dramatic speed-up, however, at the cost of computing these eigenvectors,
whose number grows with the volume.  L\"uscher realized the dominant
contribution to this space can be constructed from localized
modes\cite{algo:coherence} in a very cheap way.  Basically, very few
approximate eigenvectors of the Dirac operator are spatially cut apart
and then recombined in all possible ways. This gives then the major part
of the low eigenspace up to some physical energy. The achievement of
this method is that its cost  just grows linearly with the volume,
instead of previous methods, which had at least a $V^2$ scaling.

For the solution of the Dirac equation, removing the space constructed
from these local modes virtually eliminates all increase in the cost of
this operation as the quarks get lighter, i.e. the factor of $m^{-1}$ in
Eq.~\ref{eq:costdd}. At moderately light quark masses, gains of a factor
of 10 have been observed. These savings have been demonstrated in the
original DD-HMC setup\cite{algo:L3} as well as in mass preconditioned
HMC\cite{Marinkovic:2010eg}. A very similar idea is the adaptive
multi-grid\cite{Osborn}.

It remains to be noted, that the idea of treating the infra-red part of the 
Dirac operator's spectrum special has not stopped here. Also for observables,
computing the contribution from the infra-red more precisely than the UV part
has proven to be very beneficial\cite{DeGrand:2004qw,Giusti:2004yp}.

\section{Approaching the continuum\label{sec:cont}}
The advances in the fermionic sectors have led to significant 
optimism and large ensembles of gauge configurations at different
sea quark masses and lattice spacings have been produced. However,
going to finer lattices, a severe slowing down of the simulations 
has been observed\cite{Schaefer:2009xx,Schaefer:2010hu}.
This is not new phenomenon and had been observed previously in particular
in pure gauge theory, however, with dynamical fermions the 
problem is slightly  less severe.

\subsection{Background}
In Markov Chain Monte Carlo simulations, an algorithm is a probabilistic
procedure to generate a sequence of field configurations $U_i$
\[
U_1 \to U_2 \to U_3 \to \cdots \to U_N 
\]
given by a transition probability $T(U' \leftarrow U)$. Under certain
conditions, in particular stability
\[
P(U) = \sum_{U'} P(U') T(U\leftarrow U')
\]
the $U_i$ are then distributed according to a given probability 
distribution $P$. 
Because of this process, the probability distribution of $U_{i+1}$
depends on $U_i$, which leads to correlations among subsequent
measurements of observables $A_i=A(U_i)$. These are described by
the auto-correlation function 
\[
\Gamma_{A}(t) = \langle (A_i-\langle A \rangle)( A_{i+t}-\langle A \rangle)\rangle
\]
and in an even more concise way by the integral, the auto-correlation time
\[
\tau_\mathrm{int}(A)=\frac{1}{2} + \sum_{i=1}^\infty
\frac{\Gamma_{A}(t)}{\Gamma_{A}(0)} \ .
\]
The error $\sigma_A$ for an estimate from $N$ subsequent  measurements is then given by
\[
\sigma_A=\frac{\sqrt{\mathrm{var}(A)}}{\sqrt{N/2\tau_\mathrm{int}}} \ .
\]
This is the ordinary error formula, which effectively differs from the one
without correlation just by the reduction of the number of measurements $N$
by a factor of $2\tau_\mathrm{int}$.

\subsection{Critical slowing down}
The cost of a simulation is thus proportional to $\tau_\mathrm{int}$ and
 in the continuum limit---approaching a 
continuous phase transition---one expects it to grow with a power-law
\[
\tau_\mathrm{int}\propto a^{-z}
\]
with $a$ the lattice spacing and $z$ the {\sl dynamical} critical exponent.
The whole phenomenon is called critical slowing down and can be viewed in
analogy to static critical phenomena and their scaling laws. However, it has to
be stressed that the value of $z$ does not only depend on the properties of the
underlying theory, but also on the algorithm with which it is simulated. For
some spin models, algorithms with $z\approx0$ have been found, like cluster or
multi-grid algorithms, however, for Yang-Mills theory or  QCD, such an
algorithm does not exist. For a generic small step algorithm, one expects
$z\approx2$, based on the idea that information is distributed in a random walk
and needs to spread over a correlation length (here proportional to $a^{-1}$)
to give an independent measurement. The cost formulae in Eqs.~\ref{eq:bw} and \ref{eq:costdd}
seem to assume $z\approx 1$.

The cost of an independent measurement is  proportional
to the cost of a single update times the number of updates needed. A single update
will most certainly have a cost proportional to $a^{-4}$ for fixed volume in four
dimensions. For Hybrid Monte Carlo,  $a^{-5}$ is a typical behavior. Combined with
the effect of critical slowing down, this gives  $a^{-5-z}$, which can give 
a very sizeable  effect for a large $z$. 

\subsection{Hybrid Monte Carlo}
As described above, virtually all current simulations are using 
a variant of the HMC algorithm. It is therefore pivotal to know the behavior 
of their cost when the continuum limit is approached.
Although this cost will depend on the observable in question, a safe simulation
needs to be in a situation where all observables decorrelate much faster than
the full statistics. This is necessary, because the modes, which are being
moved by the transition matrix $T$, couple to all observables---barring some symmetry
explicitly prohibiting the coupling.
The situation needs to be such that one is able to detect from the simulation
itself the coupling of slowly moving modes to the observables in question,
which requires that we see sufficient movement in all possible
quantities.

\begin{figure}[tb]
\includegraphics[width=0.45\textwidth]{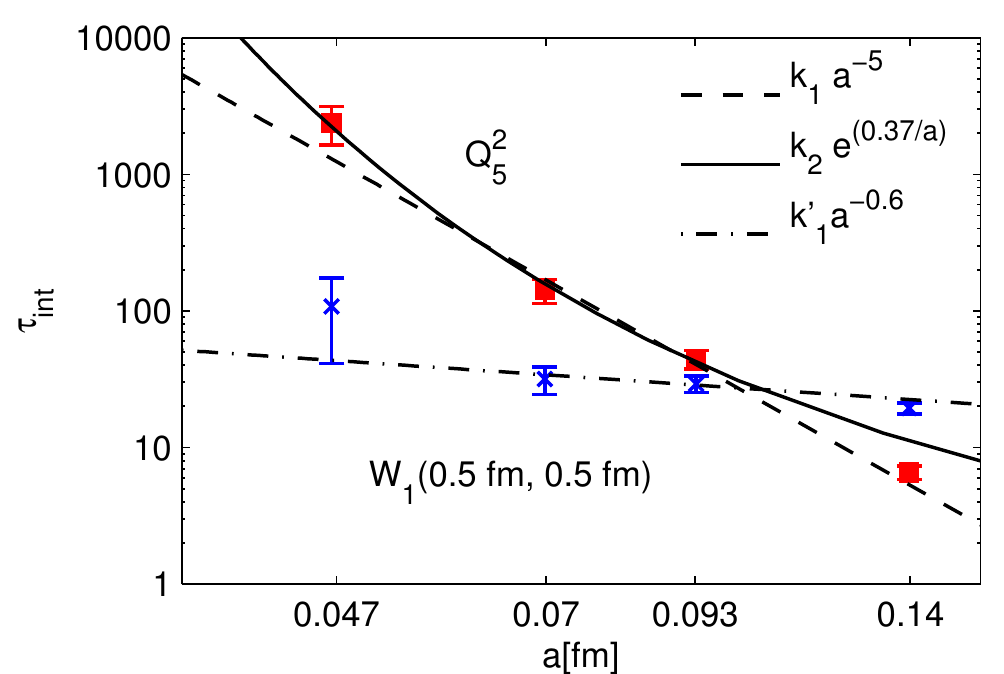}
\caption{\label{fig:1}Integrated auto-correlation time vs. the lattice
spacing. $Q^2$ is the topological charge squared, whose behavior is
compatible with $z\approx 5$, $W$ is the square Wilson loop which
for which $z\approx1$.}
\end{figure}

In order to determine the dynamical critical exponents of lattice
simulations, we performed a pure gauge study with the Wilson gauge
action  using the same algorithms
as used for QCD simulations, mainly DD-HMC, but we also tested the
behavior of pure HMC\cite{Schaefer:2010hu,Schaefer:2010aa}. The main result
is displayed in Fig.~\ref{fig:1}. It shows how the integrated
auto-correlation time rises as the continuum limit is approached. This
behavior is different for different observables. We display the
topological charge, for which a very steep rise compatible with a
dynamical critical exponent of $z\approx5$.  However, an exponential
behavior---for which evidence is presented in \cite{DelDebbio:2004xh}---
cannot be excluded.  The picture is very different for the (smeared)
square Wilson loop of size 0.5fm, which we show because it turns out to
be the loop with the slowest evolution. The rise is compatible with
$z\approx1$, actually less severe than a simple random walk picture
suggests. This is already an indication that a decoupling between the
slow modes governing the topological charge and some other observables
occurs.  The problem is that in principle, one has to check observable
by observable, whether the situation is under control, or whether the
slow modes contribute significantly.

\begin{figure}[t]
\includegraphics[width=0.30\textwidth]{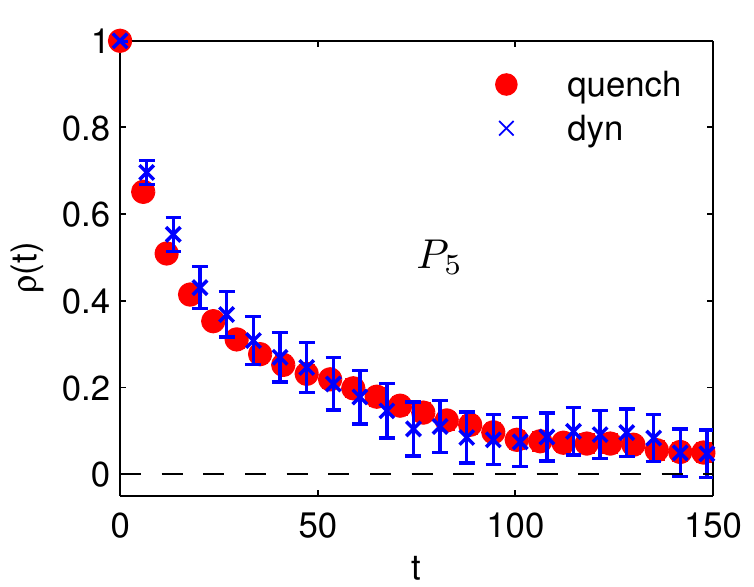}
\includegraphics[width=0.30\textwidth]{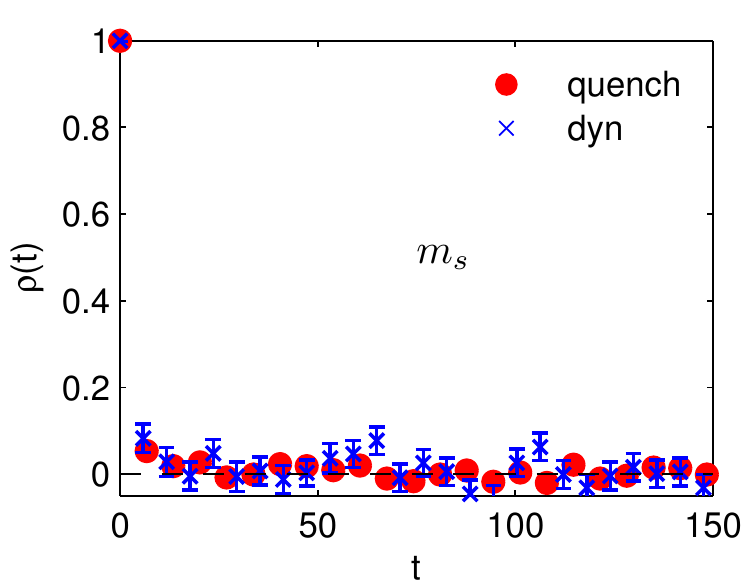}
\includegraphics[width=0.30\textwidth]{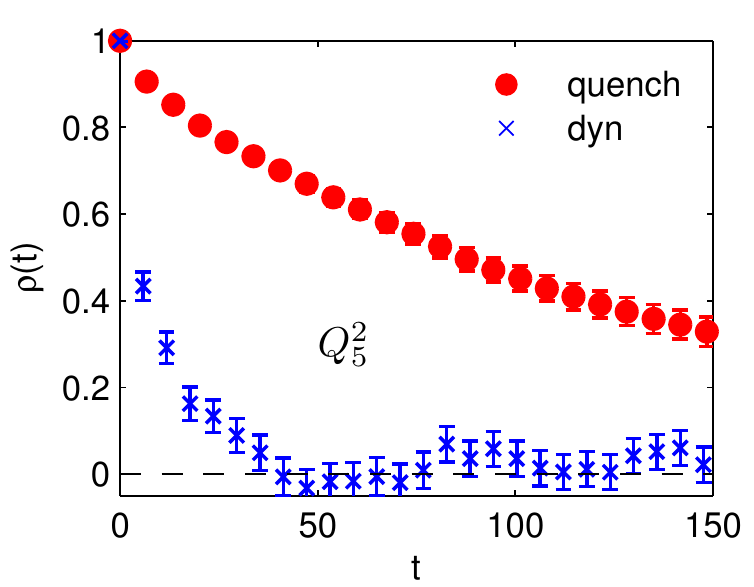}
\caption{\label{fig:2}Comparison between the normalized auto-correlation function 
$\rho(t=\Gamma(t)/\Gamma(0)$ for pure gauge and dynamical $N_f=2$ simulations. 
On the left, we show the smeared plaquette, then in the center the quark mass 
corresponding to the strange quark, on the right the square of the topological charge.}
\end{figure}

In Fig.~\ref{fig:2}, we make a comparison between the auto-correlation
behavior of pure gauge theory and of dynamical two-flavor simulations
with Wilson quarks at roughly the same lattice spacing. It is striking
that the auto-correlation functions are virtually identical between the
two set-ups for all observables at which we looked, including meson
masses, quark masses, decay constants and  the plaquette. Only in the
topological charge a significant difference appears. The same phenomenon
can also be observed in pure gauge theory, when moving to a different
gauge action. We tested it for the Iwasaki gauge action and also found the only
difference compared to the Wilson gauge action in the  auto-correlations
of the topological modes.

The slowing down of the topological charge does not come as a complete
surprise.  It has long been part of the folklore of the field that in
the continuum there is a  separation between the topological
charge sectors. From the point of view of the lattice, however, the
emergence of these sectors has long been unclear. A typical gauge
configuration is very rough and contains a large number of small objects
(dislocations) which carry topological charge.  This makes a separation
in charge sectors of the connected field space seem arbitrary.

Recently, L\"uscher\cite{Luscher:2010iy} proposed a new definition of
the charge using the gradient flow on the field space,  smoothing the
fields in a well defined way. He could show that the configurations, to
which a certain charge cannot be attributed unambiguously, die out very
quickly as the lattice spacing is reduced with a power of roughly
$a^{-6}$.  It is  clear, that a small step algorithm has problem with
this kind of situation: since the algorithm is supposed to do importance
sampling, it will stay during the evolution in the region of important
configurations; steps bridging a less important region are by
construction not included.  If the configurations between sectors are
rapidly becoming less important,  the transition from
one topological sector to the other will obviously be suppressed accordingly.

\section{Status of Wilson fermion simulations\label{sec:state}}

The ability to simulate QCD at light fermion masses and small lattice
spacings has put many collaborations into the position to generate
ensembles on which interesting physics can be studied. An overview of
the status of the simulations with Wilson fermions can be found in
Tab.~\ref{t:1}.  All collaborations use formulations, in which 
leading lattice effects are removed.
CLS\cite{vonHippel:2008pc,Brandt:2010ed} and PACS-CS\cite{Aoki:2009ix}
both use non-perturbative improvement, ETMC
\cite{Boucaud:2008xu,Baron:2010bv} uses twisted mass fermions at maximal
twist which
profit from automatic ${\cal O}(a)$ improvement. The BMW
collaboration\cite{Durr:2010vn} uses tree-level improved Wilson fermions
which couple through ``HEX'' smeared links to the gauge field.
The QCDSF and UKQCD collaborations also use gauge field smoothing
in their non-perturbatively improved SLiNC
fermions\cite{Bietenholz:2009fi}.
 All
collaborations have produced data sets with light sea quarks.  The
problems described in the previous section limit current studies to
about $0.05$fm, and even there, worries about slow modes are in place.
There are two major obstacles for even lighter fermions these days:
finite volume effects and instabilities or possible lattice phase
transitions.

The effects of finite volume can easily be kept small by a  lattice size
much larger than the pion wave length $L \gg m_\pi^{-1}$. However,
larger $L$ still comes with the fifth power in all cost formulae.
Cutting  the pion mass by half therefore requires roughly a factor of 30
in computer time, just from this criterion alone.

\begin{table}
\begin{tabular}{cccll}
\hline
collaboration \ \ \ & fermion  action    &$N_f$  &  $a$ [fm]   & $m_\pi$ [MeV] \\
\hline
BMW      & tl HEX Wilson    &2+1    & $0.05 \dots 0.12$    & 120\dots \\
CLS      & NP imp. Wilson   &2      & $0.05 \dots0.09$     & 250\dots     \\
ETMC     & tw. Wilson       &2      & $0.05 \dots 0.1$     & 280\dots  \\
         &                  &2+1+1  & $0.08 \dots 0.09$    & 270\dots   \\
PACS-CS  & NP imp. Wilson   &2+1    & 0.09                 & 135\dots  \\
\hline
\end{tabular}
\caption{Status of current Wilson fermion simulations: $N_f$ is  the number of
sea quark flavors, $a$ the lattice spacing and $m_\pi$ the
minimal sea pion mass.\label{t:1}}
\end{table}

The second obstacle are instabilities or even phase
transitions as the quark mass is lowered. Since this write-up is about
Wilson quarks, which explicitly break chiral symmetry, the latter is a
real possibility (e.g. the Aoki phase discussed in
Ref.~\cite{aokiphase}).  Instabilities in the
simulations occurring at small quark mass have been discussed\cite{DelDebbio:2005qa}, which come
from the fact that the spectrum of the Wilson operator, because of the
lack of chiral symmetry, is not bounded from below by the quark mass.

Therefore, all types of   Wilson fermions can only reach a finite
minimal quark mass for a given lattice spacing. The finer the lattice,
the smaller this minimal mass. How fine a lattice one needs to simulate
the light quarks at their physical parameters depends on the action.
The way this problem is mitigated in current set-ups used by the
PACS-CS\cite{Aoki:2009ix} or the BMW\cite{Durr:2010vn} collaboration is
to use a special gauge action or smeared fermion action, respectively,
which suppress the dislocations,  thought to be the cause of the
problem. These collaborations therefore could report simulations at the
physical value of the light quark masses.

\section{Summary}

Lattice computations have gone a long way. Light quarks are light and
their effect is taken into account in the vacuum. However, they are
still expensive. In particular, decreasing the lattice spacing comes
with a significant critical slowing down of observables
like the topological charge. The measured $z\approx 5$ means that the 
total cost of simulations rises like $a^{-10}$. 
This requires either very sizable computing
resources or a better algorithm which moves these fields more
efficiently. Such an algorithm has not been found yet. Certainly, the
phenomenon casts doubt on whether the statistical errors in current
simulations are truly under control and, with it, the extrapolation to the
continuum limit which is so crucial for final answers.

Given the decoupling between the modes which are prominent in the
topological charge and responsible for its slow evolution and, e.g. the
Wilson loop apparent in Fig.~\ref{fig:1}, one might wonder whether it
matters at all. In particular in a large volume, the {\it global}
topological charge should have very little effect on any reasonably
local observable. However, from a general Monte Carlo perspective, we can
determine the expectation values observables, their fluctuations and the
associated auto-correlations only from the simulation itself. 
Since we now know of auto-correlations of the size of a typical total 
statistics, we have to worry about even longer, undetected ones.
 To gain
confidence in any Markov Chain Monte Carlo, it is necessary that all
correlations are much smaller than the total length of the chain. To
believe that only the topological charge is affected by the increase in
auto-correlations just seems naive.

In the end, for the continuum extrapolation, the lattice simulations are
at a similar point now as they were ten years ago for the chiral
extrapolation. Back then, simulating pions with less then 500MeV seemed
impossible, with costs exploding with the sixth power. Now we are facing
a similar exponent for the increasing auto-correlations towards the
continuum. But the solution of the problem of the chiral limit gives
hope that also the continuum extrapolation will find a solution in the
not too distant future.

\paragraph{Acknowledgements}

It is a pleasure to thank M.~Hasenbusch, M.~Marinkovic, R.~Sommer and F.~Virotta for many
interesting discussions and fruitful collaboration on various subjects
which contributed to this talk. This work has been  supported in part
by the DFG through the SFB TR/9.

\end{document}